\pdfoutput=1

\documentclass[11pt,twoside,a4paper,cmspaper,final,collab]{cms-tdr}
\def\svnVersion{7322a28-D}\def\svnDate{2020/01/03}\def\cmsCernNoTag{CERN-EP-2019-114}\def\cmsCernDate{\today}\def\cmsMessage{"Published in Physics Letters B as \href{http://dx.doi.org/10.1016/j.physletb.2019.135183}{\doi{10.1016/j.physletb.2019.135183}.}"}

\begin{document}\cmsNoteHeader{SUS-18-005}

\hyphenation{had-ron-i-za-tion}
\hyphenation{cal-or-i-me-ter}
\hyphenation{de-vices}
\RCS$Revision$
\RCS$HeadURL$
\RCS$Id$

\newlength\cmsFigWidth
\ifthenelse{\boolean{cms@external}}{\setlength\cmsFigWidth{0.80\columnwidth}}{\setlength\cmsFigWidth{0.47\textwidth}}
\ifthenelse{\boolean{cms@external}}{\providecommand{\cmsLeft}{upper\xspace}}{\providecommand{\cmsLeft}{left\xspace}}
\ifthenelse{\boolean{cms@external}}{\providecommand{\cmsRight}{lower\xspace}}{\providecommand{\cmsRight}{right\xspace}}
\providecommand{\NA}{\ensuremath{\text{---}}}
\providecommand{\CL}{CL\xspace}
\newlength\cmsTabSkip\setlength{\cmsTabSkip}{1ex}
\ifthenelse{\boolean{cms@external}}{\providecommand{\cmsTable}[1]{#1}}{\providecommand{\cmsTable}[1]{\resizebox{0.95\textwidth}{!}{#1}}}
\providecommand{\cmsTableWide}[1]{\resizebox{\textwidth}{!}{#1}}
\newcommand{\mOne}{\ensuremath{{M_\mathrm{1}}}\xspace}
\newcommand{\mTwo}{\ensuremath{{M_\mathrm{2}}}\xspace}
\newcommand{\mThree}{\ensuremath{{M_\mathrm{3}}}\xspace}
\newcommand{\Diphoton}{Diphoton}
\newcommand{\PhoLep}{Photon+Lepton}
\newcommand{\PhoHTG}{Photon+\HTG\xspace}
\newcommand{\PhoST}{Photon+\ensuremath{S^{\gamma}_{\mathrm{T}}}}
\newcommand{\ST}{\ensuremath{S^{\gamma}_{\mathrm{T}}}}
\newcommand{\HTG}{\ensuremath{\HT^{\gamma}}}
\newcommand{\phopT}{\ensuremath{\pt^{\gamma}}}
\newcommand{\chizz}{\PSGczDo}
\newcommand{\chizt}{\PSGczDt}
\newcommand{\chipmpm}{\ensuremath{\PSGc_1^{\pm}}} 
\newcommand{\chipmmp}{\ensuremath{\PSGc_1^{\mp}}} 

\cmsNoteHeader{SUS-18-005}

\title{Combined search for supersymmetry with photons in proton-proton collisions at $\sqrt{s}=13\TeV$}

\date{\today}

\abstract{
A combination of four searches for new physics involving signatures with at least one photon and large missing transverse momentum, motivated by generalized models of gauge-mediated supersymmetry~(SUSY) breaking, is presented. All searches make use of proton-proton collision data at $\sqrt{s}=13\TeV$, which were recorded with the CMS detector at the LHC in 2016, and correspond to an integrated luminosity of 35.9\fbinv. Signatures with at least one photon and large missing transverse momentum are categorized into events with two isolated photons, events with a lepton and a photon, events with additional jets, and events with at least one high-energy photon. No excess of events is observed beyond expectations from standard model processes, and limits are set in the context of gauge-mediated SUSY. Compared to the individual searches, the combination extends the sensitivity to gauge-mediated SUSY in both electroweak and strong production scenarios by up to 100\GeV in neutralino and chargino masses, and yields the first CMS result combining various SUSY searches in events with photons at $\sqrt{s}=13\TeV$.
}

\hypersetup{
	pdfauthor={CMS Collaboration},
	pdftitle={Combined search for supersymmetry with photons in proton-proton collisions at sqrt(s)=13 TeV},
	pdfsubject={CMS},%
	pdfkeywords={CMS, physics, supersymmetry, gauge-mediated supersymmetry}}

\maketitle

\section{Introduction}
\label{sec:intro}

The search for supersymmetry~(SUSY), a possible theoretical extension of the standard model (SM) of particle physics, is an important piece of the physics program at the CERN LHC. Supersymmetry provides solutions to several unsolved problems in particle physics, including a mechanism for stabilizing the Higgs boson mass at the electroweak~(EW) energy scale. Supersymmetric models with a general gauge-mediated~(GGM) SUSY breaking mechanism~\cite{Dine:1981gu, AlvarezGaume:1981wy, Nappi:1982hm, Dine:1993yw, Dine:1994vc, Dine:1995ag} and $R$-parity conservation~\cite{Farrar:1978xj} often lead to final states containing photons and a large transverse momentum imbalance~\cite{Dimopoulos:1996gy, Martin:1996zb, Poppitz:1996xw, Meade:2008wd, Buican:2008ws, Abel:2009ve, Carpenter:2008wi, Dumitrescu:2010ha}. These final states are probed by several searches based on proton-proton~(pp) collisions at a center-of-mass energy~($\sqrt{s}$) of~13\TeV recorded with the ATLAS~\cite{ATLASCollaboration:2016wlb, Aaboud:2018doq} and CMS experiments~\cite{diphoton, photonlep, photonst, photonht}.

In this Letter, a combination of four different searches focusing on GGM SUSY scenarios is presented. In GGM models, the gravitino~(\sGra) is the lightest SUSY particle (LSP) and escapes undetected, leading to missing transverse momentum (\ptmiss). For these scenarios, the experimental signature depends on the nature of the next-to-LSP (NLSP), which is an admixture of the SUSY partners of EW gauge bosons. The interpretation of the combination focuses only on bino and wino, which are the superpartners of the SM U(1) and SU(2) gauge eigenstates, respectively. In most GGM models, the NLSP is assumed to be a bino- or wino-like neutralino, or a wino-like chargino. In the models used in this analysis the lightest neutralino~($\chizz$) corresponds to the NLSP, which decays to a \sGra accompanied by a photon ($\gamma$) or a \PZ~boson depending on its composition. The lightest chargino~($\chipmpm$) is assumed to decay to a \PW~boson along with a~$\chizz$ or a~\sGra. The results are interpreted in a GGM signal scenario with photons in the final state varying the bino and wino mass parameters.

To provide results for a broader set of signal topologies, the results are also interpreted in the context of simplified model scenarios~(SMS)~\cite{Chatrchyan:2013sza}. In the case of strongly produced SUSY particles, gluino and squark decays result in additional jets in the final state along with the NLSP decay products. For both EW and strong SUSY production, the gaugino branching fractions are varied to probe a range of possible scenarios resulting in final states with photons, \PZ~or \PW~bosons.

All searches used in the combination are performed with pp collision data at~$\sqrt{s}=13\TeV$, corresponding to an integrated luminosity of 35.9\fbinv, collected with the CMS detector in 2016. In the combination, each search corresponds to a category of events. The first category requires the presence of two isolated photons (\Diphoton{} category). This category is based on the search presented in Ref.~\cite{diphoton} and targets bino-like neutralino decays. Events with electrons~($\Pepm$) or muons~($\mu^\pm$) are vetoed in this category. The \PhoLep{} category requires one isolated photon, as well as one isolated $\Pepm$ or $\mu^\pm$. This category is based on the search presented in Ref.~\cite{photonlep} and targets wino-like chargino decays along with bino-like neutralino decays. The \PhoST{} category requires the presence of at least one isolated photon and large \ptmiss utilizing the variable $\ST=\ptmiss + \sum_{\gamma_i} \pt^{\gamma_i}$, where $\pt^{\gamma_i}$ is the transverse momentum of photons in the event. This search, presented in Ref.~\cite{photonst}, provides sensitivity to both EW and strong production. The \PhoHTG{} category is based on the search presented in Ref.~\cite{photonht} and focuses on strongly produced gluinos and squarks. This search requires at least one isolated photon and significant hadronic activity by selecting events with large values of $\HTG=\HT+ \phopT$, where $\HT$ is the scalar sum of all jet momenta and $\phopT$ is the transverse momentum of the leading photon in the event.

To ensure exclusive search regions for the combination, any overlapping kinematic regions in the four categories are combined such that a single event is only present in one category. For SUSY scenarios that are based on EW production, all four categories are used. For strong SUSY production, the \Diphoton{} category is removed.

\section{Signal scenarios}
\label{sec:ggm}
\begin{figure*}[h!]
	\centering
	\includegraphics[width=\cmsFigWidth]{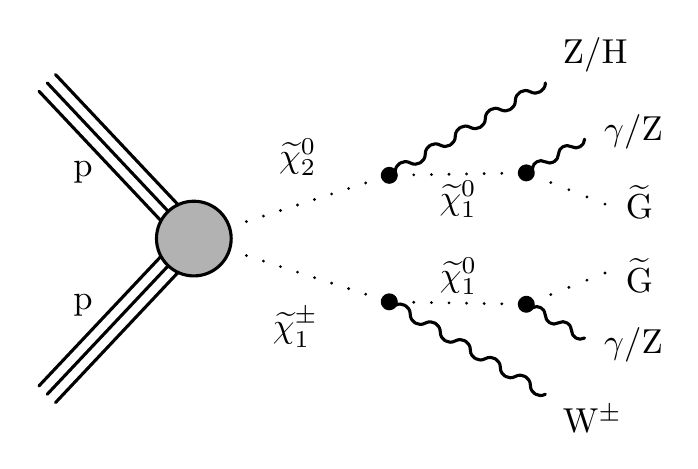}\hfil
	\includegraphics[width=\cmsFigWidth]{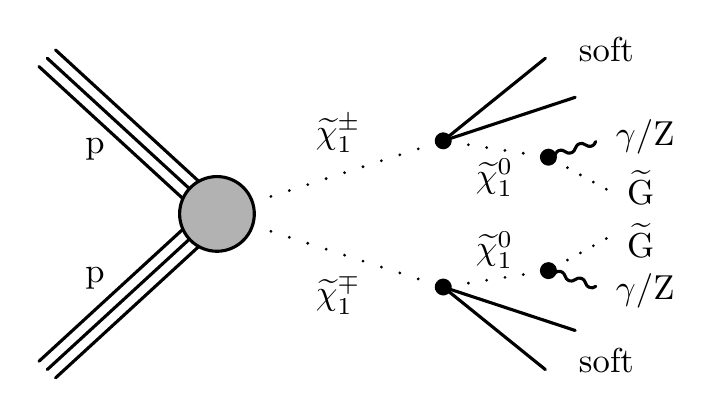}\\
	\null
	\includegraphics[width=\cmsFigWidth]{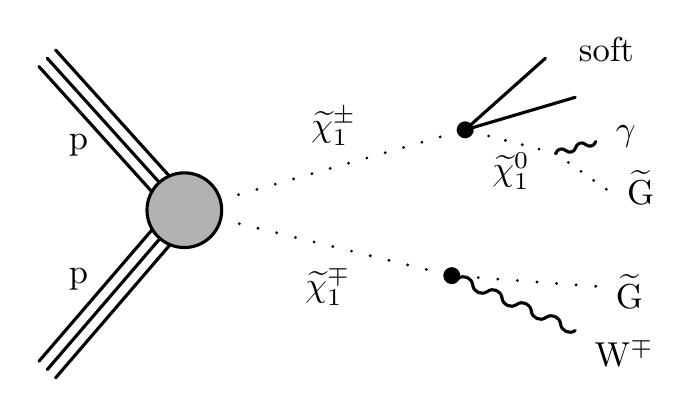}\hfil
	\includegraphics[width=\cmsFigWidth]{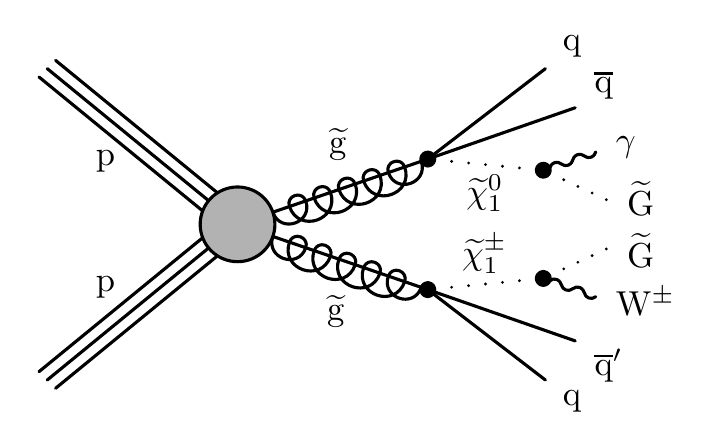}
	\caption{Diagrams of the SUSY processes considered in this Letter: one process within the GGM scenario~(upper left), two EW SMS processes, with possible neutralino and chargino decays~(upper right and lower left), and a strong SMS process based on gluino pair production~(lower right).}
	\label{fig:GGM}
\end{figure*}

The SUSY scenarios considered in this Letter are sketched in Fig.~\ref{fig:GGM}; they include one GGM scenario~(upper left), two EW SMS~(upper right and lower left), and one strong production SMS~(lower right).

For the GGM scenario, the squark and gluino masses are set to be large, rendering them irrelevant to the studied LHC collisions and ensuring that strong production is negligible and EW production of gauginos, namely $\chipmpm\chipmmp$, $\chipmpm\chizz$, and $\chipmpm\chizt$ production, is dominant. The GGM framework used to derive the GGM scenario is suitable for unifying models of gauge-mediation in a more general way with only a few free parameters~\cite{Grajek:2013ola,Knapen:2016exe,Knapen:2015qba}. For the GGM scenario considered in this Letter, the techniques of Ref.~\cite{Knapen:2016exe} are used to reduce the 8-dimensional GGM parameter space to two gaugino mass parameters. The GGM scenario is defined by setting the GGM parameters as follows:
\begin{align*}
	\mThree &= \mu=8\TeV,\\
	m_{\mathrm{Q}} &= m_{\mathrm{U}}=10\TeV,\\
	m_{\mathrm{D}} &= 8\TeV.
\end{align*}
All parameters are defined at the messenger scale, which is set to $M_\mathrm{mess}=10^{15}\GeV$.
The parameters \mThree and $\mu$ are the gluino and higgsino mass parameters, respectively, and the parameters $m_{\mathrm{Q}}$, $m_{\mathrm{U}}$, and $m_{\mathrm{D}}$ are the sfermion soft masses. In this GGM scenario, the remaining bino~($\mOne$) and wino~($\mTwo$) mass parameters are varied and the Higgs boson mass receives large radiative corrections from the heavy stops to yield the observed mass at the EW scale.

In GGM, the lifetime of the NLSP is a function of the NLSP and the gravitino masses. In order to ensure prompt decays of the NLSP in the detector, the gravitino mass is fixed to 10\unit{eV}. As was shown in Ref.~\cite{Knapen:2015qba}, this implies heavy squarks ($m_{\PSQ}\gtrsim3\TeV$), which is consistent with the model used in this Letter.

\begin{figure}[tbh!]
	\centering
	\includegraphics[width=0.49\textwidth]{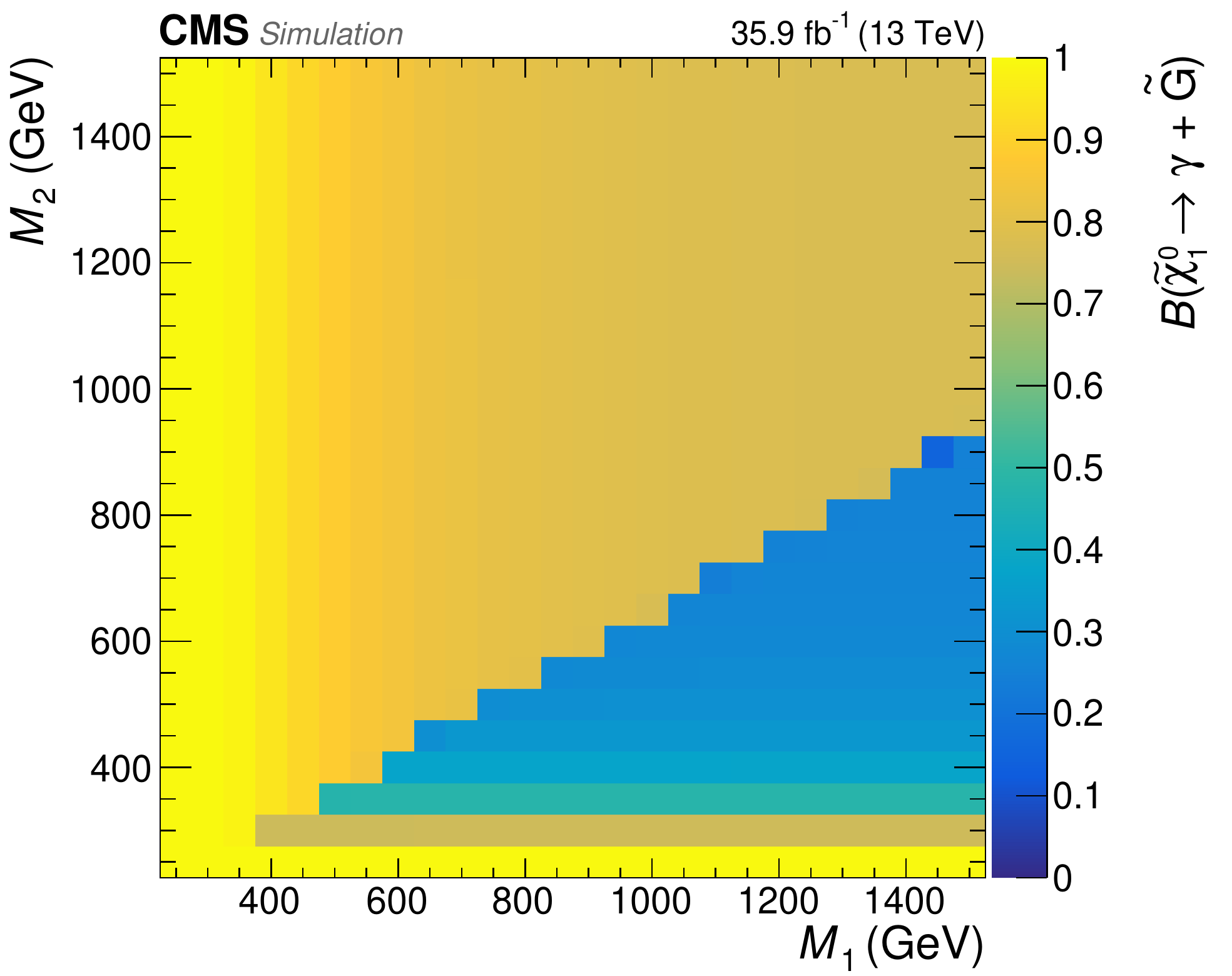}\hfil
	\includegraphics[width=0.49\textwidth]{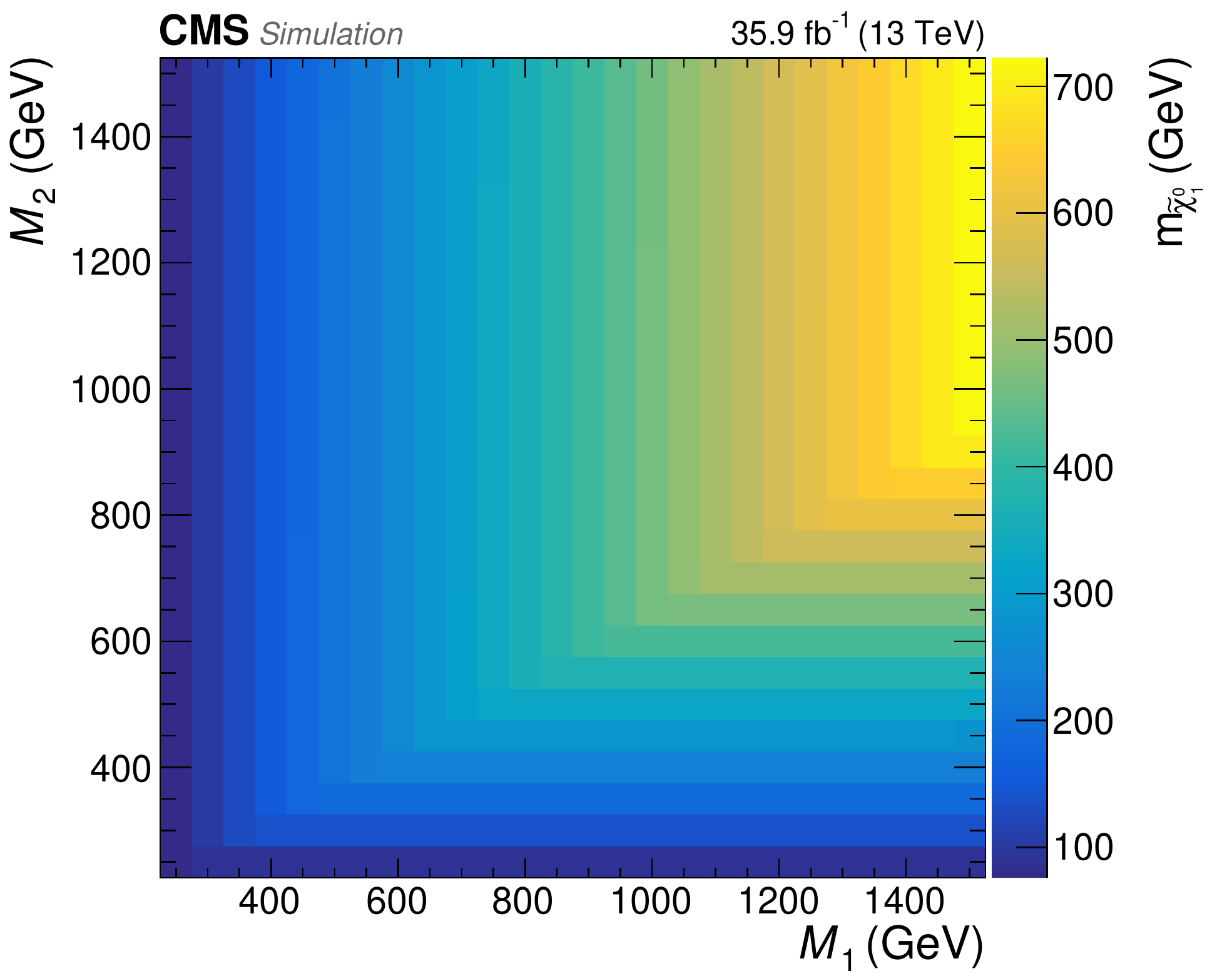}\\
	\caption{Branching fractions~(\textit{B}) for the NLSP decay to a photon and a gravitino for the GGM scenario (\cmsLeft). The phase space is spanned by the bino~($\mOne$) and wino~($\mTwo$) mass parameters showing the change of the NLSP composition. This change also influences the dependence of the physical mass of the neutralino~($m_{\chizz}$) on the gauge mass parameters (\cmsRight).}
	\label{fig:BF}
\end{figure}

One possible diagram for the GGM scenario is shown in Fig.~\ref{fig:GGM}~(upper left). The chargino always decays to the \PW~boson along with the lightest neutralino, and the $\chizt$ could decay to a \PZ~boson or an \PH boson along with the lightest neutralino. The branching fraction of the NLSP decaying into a photon and a gravitino is determined by the composition of the gauge eigenstates of the NLSP. As shown in Fig.~\ref{fig:BF} (\cmsLeft), the branching fraction of the NLSP changes across the parameter space. For large $\mOne$ and medium $\mTwo$, the NLSP is wino-like. This increases the branching fraction for $\chizz\to\PZ\sGra$ decays in the phase space of $\mTwo\gtrsim300\GeV$ where the NLSP mass exceeds the \PZ~boson mass. In the remaining phase space, the NLSP is bino-like, which increases the $\chizz\to\gamma\sGra$ branching fraction. The different compositions of the NLSP can also be extracted from the dependence of the physical NLSP mass on the model parameters $\mOne$ and $\mTwo$, as shown in Fig.~\ref{fig:BF}~(\cmsRight). With a wino-like NLSP, the physical mass scales with $\mTwo$, whereas, for the remaining phase space with bino-like NLSPs, the physical mass depends on $\mOne$.

Based on EW production SMSs, two different branching fraction scenarios are constructed. For these scenarios, the chargino and neutralino masses are almost degenerate in mass, such that the \PW~boson from the chargino decay is produced off-shell, resulting in low momentum~(soft) particles that are outside the detector acceptance. In the case of the neutralino branching fraction scenario, $\chipmpm\chizz$ and $\chipmpm\chipmmp$ are probed as shown in Fig.~\ref{fig:GGM}~(upper right). In this scenario, the chargino always decays to the NLSP, whereas the branching fractions for the decay modes $\chizz\to\gamma\sGra$ and $\chizz\to \PZ\sGra$ are varied. In the chargino branching fraction scenario, shown in Fig.~\ref{fig:GGM}~(lower left), the chargino can decay to the LSP or NLSP, and the branching fractions for the decay modes $\chipmpm\to \PW^{\pm}\sGra$ and $\chipmpm\to\chizz(+\mathrm{soft})$ are varied. The decay mode of $\chizz\to\gamma\sGra$ is fixed. In this scenario only $\chipmpm\chipmmp$ is produced. These scenarios probe a range of NLSP compositions with bino- and wino-like neutralinos, and wino-like charginos.

The strong production SMS, shown in Fig.~\ref{fig:GGM}~(lower right), is used for the nominal gluino scenario and the gluino branching fraction scenario. In both scenarios, gluino pair production is probed, assuming the gluino decays to a chargino or neutralino. The decay modes for the neutralino and chargino, which are assumed to be mass degenerate, are fixed to $\chizz\to\gamma\sGra$ and $\chipmpm\to \PW^{\pm}\sGra$, respectively. In the nominal gluino scenario, the gluino branching fractions to either charginos or neutralinos are both set to 50\%, and the gluino and NLSP masses are varied. Only light flavor quarks, udsc, are included from the gluino decay. This probes a range of scenarios where the gluino mass is small. In the gluino branching fraction scenario, the gluino branching fractions are varied along with the NLSP mass, and the gluino mass is set to 1950\GeV, which corresponds to the gluino mass where the largest gain from the combination is expected.

The production cross sections for all points in the GGM scenario are computed at next-to-leading order~(NLO) using the \PROSPINO{2} framework~\cite{Beenakker:1996ed}. The uncertainties in the cross section calculation are derived with \PROSPINO{2} following the PDF4LHC recommendations~\cite{Butterworth:2015oua} and using the parton distribution functions~(PDFs) in the LHAPDF data format~\cite{Buckley:2014ana}. The simulation incorporates the NNPDF 3.0~\cite{Ball:2014uwa} PDFs and uses \PYTHIA{8}~\cite{Sjostrand:2007gs} with the CUETP8M1 generator tune to describe parton showering and the hadronization~\cite{Khachatryan:2015pea}. The simplified model signals are generated with \MGvATNLO 2.2.2, including up to two additional partons, at leading order~\cite{Alwall:2014hca} and scaled to NLO and NLO~+~next-to-leading logarithmic~ accuracy~\cite{Beenakker:1999xh,Fuks:2012qx,Fuks:2013vua,Beenakker:1996ch,Kulesza:2008jb,Kulesza:2009kq,Beenakker:2009ha,Beenakker:2011fu,Borschensky:2014cia}. All generated signal events are processed with a fast simulation of the CMS detector response~\cite{CMS:2010spa}. Scale factors are applied to compensate for any differences with respect to the full simulation, which is based on \GEANTfour~\cite{Agostinelli:2002hh}.

\section{The CMS detector}
\label{sec:CMS}

The central feature of the CMS apparatus is a superconducting solenoid of 6\unit{m} internal diameter, providing a magnetic field of 3.8\unit{T}. Within the solenoid volume are a silicon pixel and strip tracker, a lead tungstate crystal electromagnetic calorimeter (ECAL), and a brass and scintillator hadron calorimeter (HCAL), each composed of a barrel and two endcap sections. Forward calorimeters extend the pseudorapidity~($\eta$) coverage provided by the barrel and endcap detectors. Muons are detected in gas-ionization chambers embedded in the steel flux-return yoke outside the solenoid.

The analysis only utilizes photons measured in the barrel section of the ECAL~($\abs{\eta}<1.44$). In this section, an energy resolution of about 1\% is achieved for unconverted or late-converting photons in the tens of \GeV energy range. The remaining barrel photons have a resolution of about 1.3\% up to a pseudorapidity of $\abs{\eta} = 1.0$, rising to about 2.5\% at $\abs{\eta} = 1.4$~\cite{CMS:EGM-14-001}.

A more detailed description of the CMS detector, together with a definition of the coordinate system used and the relevant kinematic variables, can be found in Ref.~\cite{Chatrchyan:2008zzk}.

\section{Object reconstruction and identification}
\label{sec:Reco}

Photons, electrons, muons, and jets are reconstructed with the particle-flow (PF) event algorithm~\cite{Sirunyan:2017ulk}, which identifies particles produced in a collision
combining information from all detector subsystems. The energy of photons is directly obtained from the ECAL measurement. Likewise, the energy of electrons is derived from a combination of the momentum measured in the tracker and the energy measured from spatially compatible clusters of energy deposits in the ECAL. The energy of muons is obtained from the curvature of the corresponding track. ECAL and HCAL energy deposits associated to tracks are reconstructed as charged hadrons; remaining energy deposits are reconstructed as neutral hadrons. Jets are reconstructed from PF candidates using the anti-\kt clustering algorithm~\cite{Cacciari:2008gp} with a distance parameter of 0.4.

The missing transverse momentum vector \ptvecmiss is computed as the negative vector sum of the transverse momenta of all the PF candidates in an event, and its magnitude is denoted as \ptmiss. The \ptvecmiss is modified to account for corrections to the energy scale of the reconstructed jets in the event~\cite{CMS-PAS-JME-17-001}.

Photons considered in this Letter are required to be isolated and have an ECAL shower shape consistent with a single photon shower. The photon isolation is determined by computing the transverse energy of all PF charged hadrons, neutral hadrons, and other photons in a cone centered around the photon momentum vector. The cone has an outer radius of $0.3$ in $\Delta R = \sqrt{\smash[b]{(\Delta\phi)^2 + (\Delta\eta)^2}}$~(where $\phi$ is azimuthal angle in radians). The contribution of the photon to this cone is removed. Corrections for the effects of multiple interactions in the same or adjacent bunch crossing (pileup) are applied to all isolation energies, depending on the $\eta$ of the photon. The Diphoton category~\cite{diphoton} uses photon identification criteria to preserve an average photon selection efficiency of 80$\%$ while suppressing backgrounds from quantum chromodynamics~(QCD) multijet events. The other three categories~\cite{photonlep, photonst, photonht} use looser identification criteria to preserve a high photon selection efficiency of 90\%. Only photons reconstructed in the barrel region~($\abs{\eta}<1.44$) are used, because the SUSY signal models considered in this combination produce photons primarily in the central region of the detector.

Reconstructed jets are used to compute the $\HT$ variable as well as the $\HTG$ variable along with the selected photons. Jets reconstructed within a cone of $\Delta R <0.4$ around the leading photon are not considered in both variables. Jets with $\pt>30\GeV$ and $\abs{\eta}<3.0$ are used. In case of the \PhoLep{} only jets with $\abs{\eta}<2.5$ are taken into account. The \Diphoton{} category makes use of no jet variables.

Identification of electrons is based on the shower shape of the ECAL cluster, the HCAL-to-ECAL energy ratio, the geometric matching between the cluster and the track, the quality of the track reconstruction, and the isolation variable. The isolation variable is calculated from the transverse momenta of photons, charged hadrons, and neutral hadrons within a cone whose radius is variable depending on the electron \pt~\cite{Rehermann:2010vq}, and which is corrected for the effects of pileup~\cite{Cacciari:2007fd}. Hits in the pixel detector are used to distinguish electrons from converted photons.

A set of muon identification criteria, based on the goodness of the global muon track fit and the quality of the muon reconstruction, is applied to select the muon candidates. Muons are also required to be isolated from other objects in the event using a similar isolation variable as in the electron identification.

\section{Event selection}
\label{sec:EventSelection}
Events are divided into the four categories shown in Table~\ref{tab:searchbins}. Each category is based on one of the four individual searches~\cite{photonht,photonst,photonlep,diphoton}. The minimum photon \pt is mainly determined by the trigger requirements in each of the four searches. The \PhoST{} and \PhoHTG{} categories are also referred to as inclusive categories. The signal regions for these categories are defined by \ST{} and \HTG{} respectively, where the \PhoHTG{} category also has search regions in \ptmiss{}. Selected diphoton events are classified by values of \ptmiss, whereas events with a photon and a lepton are classified by values of \phopT, \HT, and \ptmiss.

\begin{table*}[htb!]
	\centering
	\topcaption{Definitions of the four exclusive categories. The kinematic selections and the search bins are based on the four individual searches, while the additional vetoes shown in the third columns ensure exclusive event categories. The transverse mass of a photon/lepton and \ptmiss is denoted as~$\mT\left(\gamma/\ell, \ptmiss\right)$. The search bins always include the lower bounds. The Diphoton and Lepton veto match the kinematic selections of the \Diphoton{} and \PhoLep{} category, respectively. The Diphoton veto is only used in the interpretation of the EW produced scenarios, but dropped for the strong produced scenarios, where the Diphoton category is not part of the combination.}
	\label{tab:searchbins}
	\renewcommand{\arraystretch}{1.2}
	\cmsTable{
		\begin{tabular}{rlcc}
			\hline
			\multicolumn{2}{c}{Kinematic selections}                                                                                            & Search bins (\GeVns)                           & Vetoed events                     \\
			\hline
			\multicolumn{4}{c}{\Diphoton{} category}                                                                                                                                                             \\
			$\phopT$&$>40\GeV$                                                                               &       & \multirow{4}{*}{\NA}              \\
			$\ptmiss$&$>100\GeV$						&\ptmiss: [100, 115], [115,130], [130,150],\\
			$m_{\gamma\gamma}$&$>105\GeV$                                                  & [150,185], [185,250], $\geq$250                    &                                   \\
			Lepton veto for $\pt^{\ell}$&$>25\GeV$ \\ [\cmsTabSkip]
			\multicolumn{4}{c}{\PhoLep{} category}                                                                                                                                                               \\
			$\phopT$&$>35\GeV$                                                            & \multirow{2}{*}{$\ptmiss$: [120, 200], [200, 400], $\geq$400}      & \multirow{4}{*}{\NA}              \\
			$\ptmiss$&$>120\GeV$ 				&\multirow{2}{*}{$\HT$: [0,100], [100, 400], $\geq$400}	\\
			$\pt^{\ell}$&$>25\GeV$                                  & \multirow{2}{*}{$\phopT$: [35,200], $\geq$200} &                                   \\
			$\mT\left(\ell, \ptmiss\right)$&$>100\GeV$                                                                                                                &                      & \\[\cmsTabSkip]
			\multicolumn{4}{c}{\PhoST{} category}                                                                                                                                                                \\
			$\phopT$ &$>180\GeV$                                                                            &                  &  \\
			$\ptmiss$&$>300\GeV$										& $\ST$: [600, 800], [800, 1000],                         & $\HTG>2\TeV$ if $\ptmiss>350\GeV$ \\
			$\ST$&$>600\GeV$           									& [1000,1300], $\geq$1300 &Diphoton, Lepton \\
			$\mT\left(\gamma, \ptmiss\right)$&$>300\GeV$													\\[\cmsTabSkip]
			\multicolumn{4}{c}{\PhoHTG{} category}                                                                                                                                                               \\
			$\phopT$&$>100\GeV$                                                                              & &                     \\
			$\ptmiss$&$>350\GeV$										&$\ptmiss$: [350,450], [450, 600], $\geq$600     &$\HTG<2\TeV$   	\\
			$\HTG$&$>700\GeV$  & $\HTG$: [700, 2000], $\geq$2000                   & Diphoton, Lepton                  \\
			$\abs{\Delta\phi\left(\pm\ptvecmiss,\vec{p}_{\mathrm{T}}^{\gamma}\right)}$&$>0.3$ \\
			\hline
		\end{tabular}
	}
\end{table*}

To enable a statistical combination of the four categories, the overlap between the categories is removed by applying additional vetoes. Since the \Diphoton{} and \PhoLep{} category show the highest sensitivities for the GGM scenario, these categories remain unchanged with respect to the initial searches. Events with leptons or two photons that are selected in the other two categories, but also match the requirements of the \Diphoton{} or \PhoLep{} categories,  are vetoed in the \PhoST{} and \PhoHTG{} categories. To remove the overlap between the two inclusive categories, the two categories are separated as follows. Events with a large hadronic activity~($\HTG>2\TeV$) are vetoed from the \PhoST{} category if they match the $\ptmiss$~requirement of the \PhoHTG{} category. In addition, events with lower hadronic activity~($\HTG<2\TeV$) are vetoed from the \PhoHTG{} category and assigned to \PhoST{}. To further increase the sensitivity to strong production the veto strategy is slightly changed for the interpretation of the gluino scenarios. For these scenarios the Diphoton category is not included in the combination and events with two photons are kept in the \PhoST{} and \PhoHTG{} categories, which have larger sensitivity to strong production.

The SM background in the \PhoST{} and \PhoLep{} categories is dominated by vector boson production with initial-state photon radiation, denoted as ``Vector-boson~+~$\gamma{}$'', which is in each case estimated from simulation scaled in a particular control region~\cite{photonst,photonlep}. For the \PhoHTG{} category, on the other hand, the $\HTG$ requirement implies hadronic activity leading to a dominant background from QCD multijet and $\gamma$~+~jet processes, which also holds for the \Diphoton{} category. In both categories data-driven methods are used to estimate this background contribution~\cite{photonht,diphoton}. Additional contributions arise from electrons which are misidentified as photons and jets which are misidentified as leptons. For both of these processes data-driven methods are utilized to estimate the contribution to the search regions. Furthermore, $\text{t}\bar{\text{t}}\gamma$ and diboson processes, summarized as ``Rare Backgrounds'', can contribute to all four categories and are estimated using simulation.

\section{Results}
\label{sec:results}

\begin{figure*}[tbh]
	\centering
	\includegraphics[width=0.89\textwidth]{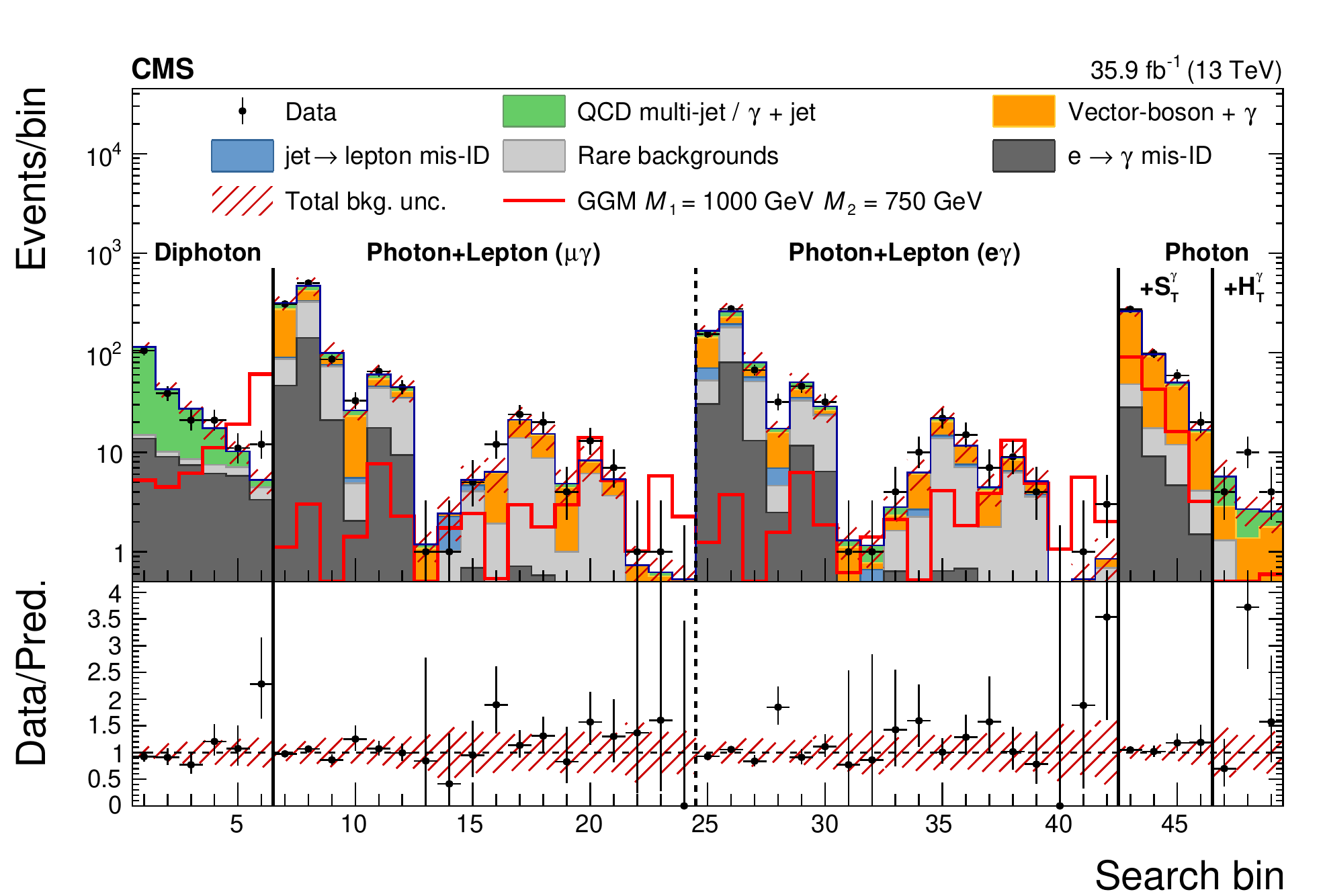}
	\caption{Predicted pre-fit background yields, where the values are not constrained by the likelihood fit, and observed number of events in data for all search bins used in the combination. The search bins are defined in Table~\ref{tab:yields}. The hatched red bands in both parts of the plot represent the total uncertainty of the background prediction. The red line in the upper panel shows the signal prediction for one specific signal point of the GGM scenario with $M_{1}=1000\GeV$ and $M_{2}=750\GeV$. The lower panel shows the ratio between the observed data and the predicted backgrounds.}
	\label{fig:results}
\end{figure*}

Figure~\ref{fig:results} and Table~\ref{tab:yields} show a comparison between the data and the background prediction for the search
bins used in the combination. In case of the \PhoLep{} and the \Diphoton{} categories, the yields correspond to the results of the published searches. The yields of the \PhoST{} and \PhoHTG{}
categories are based on the modified event selections, which ensure exclusive signal regions. Overall agreement between the observed number of events and the background prediction is found for the 49 search bins.

\begin{table*}[htbp!]
	\topcaption{Predicted pre-fit background yields, where the values are not constrained by the likelihood fit, the observed number of events in data, and the post-fit background yields after the constraint from the likelihood fit
 for all search bins used in the combination. In addition the range covered by each individual bin is shown.}
	\label{tab:yields}
	\centering
	\renewcommand{\arraystretch}{1.1}
	\cmsTableWide{
		\centering
		\begin{tabular}{c|c|c|c|c|c|c|c}
			Bin & Category                                  & \multicolumn{3}{c|}{Ranges (\GeVns)}   & Total bkg.                         & Data      & Post-fit Total bkg.                                          \\ \hline
			1   & \multirow{6}{*}{\Diphoton}                & \multirow{6}{*}{$\phopT\geq40$}          &\multicolumn{2}{c|}{$110\leq\ptmiss<115$} & $114\pm    13$                     & $       105$   & $110 \pm 9$                                    \\
			2   &                                           &                                           & \multicolumn{2}{c|}{$115\leq\ptmiss<130$} & $42.9\pm     7.2$                  & $        39$   & $41.6 \pm 5.5$                                    \\
			3   &                                           &                                           & \multicolumn{2}{c|}{$130\leq\ptmiss<150$} & $27.3\pm     5.4$                  & $        21$   & $25.9 \pm 3.6$                                    \\
			4   &                                           &                                           & \multicolumn{2}{c|}{$150\leq\ptmiss<185$} & $17.4\pm     3.9$                  & $        21$   &  $18.0 \pm 3.0$                                    \\
			5   &                                           &                                           & \multicolumn{2}{c|}{$185\leq\ptmiss<250$} & $10.2\pm     2.6$                  & $        11$   & $10.8 \pm 2.0$                                    \\
			6   &                                           &                                           & \multicolumn{2}{c|}{$\ptmiss\geq250$}     & $5.3\pm     1.4$                   & $        12$   & $5.9 \pm 1.4$                                   \\
			\hline
			7   & \multirow{18}{*}{\PhoLep{} ($\mu\gamma$)} & \multirow{9}{*}{$35\leq\phopT<200$}          & \multirow{3}{*}{$120\leq\ptmiss<200$} & $\HT<100$     & $317\pm    50$      & $       309$ & $318 \pm 19$ \\
			8   &                                           &                                        &                                    & $100\leq\HT<400$ & $470\pm    98$      & $       501$  & $490 \pm 32$\\
			9   &                                           &                                        &                                    & $\HT\geq400$     & $100\pm      27$    & $        86$ & $99 \pm 7$ \\
			\cline{4-8}
			10  &                                           &                                        & \multirow{3}{*}{$200\leq\ptmiss<400$} & $\HT<100$     & $26.3\pm     5.3$   & $        33$ & $30.5 \pm 3.1$\\
			11  &                                           &                                        &                                    & $100\leq\HT<400$ & $61\pm    14$       & $        65$ & $63 \pm 5$\\
			12  &                                           &                                        &                                    & $\HT\geq400$     & $45\pm    14$       & $        45$ & $46 \pm 5$\\
			\cline{4-8}
			13  &                                           &                                        & \multirow{3}{*}{$\ptmiss\geq400$}     & $\HT<100$     & $1.2\pm     0.4$    & $         1$ & $1.3 \pm 0.3$\\
			14  &                                           &                                        &                                    & $100\leq\HT<400$ & $2.4\pm     1.1$    & $         1$ & $2.1 \pm 0.7$\\
			15  &                                           &                                        &                                    & $\HT\geq400$     & $5.3\pm       2.0$  & $         5$ & $5.4 \pm 1.1$\\
			\cline{3-8}
			16  &                                           & \multirow{9}{*}{$\phopT\geq200$}          & \multirow{3}{*}{$120\leq\ptmiss<200$} & $\HT<100$     & $6.3\pm     2.4$    & $        12$ & $9.1 \pm 1.6$ \\
			17  &                                           &                                        &                                    & $100\leq\HT<400$ & $21.1\pm     7.2$   & $        24$ & $23.2 \pm 2.5$\\
			18  &                                           &                                        &                                    & $\HT\geq400$     & $15.3\pm     4.8$   & $        20$ & $17.2 \pm 2.0$\\
			\cline{4-8}
			19  &                                           &                                        & \multirow{3}{*}{$200\leq\ptmiss<400$} & $\HT<100$     & $4.8\pm     1.8$    & $         4$ & $6.9 \pm 1.2$\\
			20  &                                           &                                        &                                    & $100\leq\HT<400$ & $8.3\pm     3.2$    & $        13$ & $9.0 \pm 1.1$\\
			21  &                                           &                                        &                                    & $\HT\geq400$     & $5.4\pm       2.0$  & $         7$ & $6.3 \pm 0.9$\\
			\cline{4-8}
			22  &                                           &                                        & \multirow{3}{*}{$\ptmiss\geq400$}     & $\HT<100$     & $0.7\pm     0.4$    & $         1$ & $1.2 \pm 0.3$\\
			23  &                                           &                                        &                                    & $100\leq\HT<400$ & $0.6\pm     0.2$    & $         1$ & $0.8 \pm 0.1$\\
			24  &                                           &                                        &                                    & $\HT\geq400$     & $0.5\pm     0.2$    & $         0$ & $0.6 \pm 0.1$\\
			\hline
			25  & \multirow{18}{*}{\PhoLep{} $(\Pe\gamma)$}   & \multirow{9}{*}{$35\leq\phopT<200$}          & \multirow{3}{*}{$120\leq\ptmiss<200$} & $\HT<100$     & $166\pm    22$      & $       154$ & $167 \pm 12$\\
			26  &                                           &                                        &                                    & $100\leq\HT<400$ & $261\pm    53$      & $       276$ & $271 \pm 18$\\
			27  &                                           &                                        &                                    & $\HT\geq400$     & $80\pm    21$       & $        67$ & $80 \pm 7$\\
			\cline{4-8}
			28  &                                           &                                        & \multirow{3}{*}{$200\leq\ptmiss<400$} & $\HT<100$     & $17.3\pm     3.2$   & $        32$ & $21.2 \pm 2.7$ \\
			29  &                                           &                                        &                                    & $100\leq\HT<400$ & $51\pm    12$       & $        46$ & $51 \pm 4$\\
			30  &                                           &                                        &                                    & $\HT\geq400$     & $28.8\pm       9.0$ & $        32$ & $29.6 \pm 3.0$\\
			\cline{4-8}
			31  &                                           &                                        & \multirow{3}{*}{$\ptmiss\geq400$}     & $\HT<100$     & $1.3\pm     0.5$    & $         1$ & $1.5 \pm 0.4$\\
			32  &                                           &                                        &                                    & $100\leq\HT<400$ & $1.2\pm     0.5$    & $         1$ & $1.2 \pm 0.4$\\
			33  &                                           &                                        &                                    & $\HT\geq400$     & $2.8\pm     0.8$    & $         4$ & $3.3 \pm 0.6$\\
			\cline{3-8}
			34  &                                           & \multirow{9}{*}{$\phopT\geq200$}          & \multirow{3}{*}{$120\leq\ptmiss<200$} & $\HT<100$     & $6.3\pm     2.1$    & $        10$ & $8.3 \pm 1.4$ \\
			35  &                                           &                                        &                                    & $100\leq\HT<400$ & $21.8\pm     7.1$   & $        22$ & $23.3 \pm 2.4$\\
			36  &                                           &                                        &                                    & $\HT\geq400$     & $11.7\pm     3.7$   & $        15$ & $13.0 \pm 1.5$\\
			\cline{4-8}
			37  &                                           &                                        & \multirow{3}{*}{$200\leq\ptmiss<400$} & $\HT<100$     & $4.4\pm     1.6$    & $         7$ & $5.8 \pm 1.0$ \\
			38  &                                           &                                        &                                    & $100\leq\HT<400$ & $8.9\pm     3.3$    & $         9$ & $9.2 \pm 1.3$\\
			39  &                                           &                                        &                                    & $\HT\geq400$     & $5.1\pm     1.8$    & $         4$ & $5.6 \pm 0.8$\\
			\cline{4-8}
			40  &                                           &                                        & \multirow{3}{*}{$\ptmiss\geq400$}     & $\HT<100$     & $0.4\pm     0.2$    & $         0$ & $0.6 \pm 0.2$\\
			41  &                                           &                                        &                                    & $100\leq\HT<400$ & $0.5\pm     0.2$    & $         1$ & $0.7 \pm 0.2$\\
			42  &                                           &                                        &                                    & $\HT\geq400$     & $0.8\pm     0.5$    & $         3$ & $1.1 \pm 0.4$\\
			\hline
			43  & \multirow{4}{*}{\PhoST{}}                 & \multirow{4}{*}{$\phopT\geq180$}       & \multirow{4}{*}{$\HTG\leq2000$}           & \multicolumn{1}{c|}{$600\leq\ST<800$}     & $260\pm    30$                     & $       273$    &   $274 \pm 22$                                \\
			44  &                                           &                                        &                                           & \multicolumn{1}{c|}{$800\leq\ST<1000$}    & $96\pm    14$                      & $       98$      &  $100 \pm 9$                                \\
			45  &                                           &                                        &                                           & \multicolumn{1}{c|}{$1000\leq\ST<1300$}   & $50.0\pm     7.9$                  & $       59$       &  $53.8 \pm 6.4$                               \\
			46  &                                           &                                        &                                           & \multicolumn{1}{c|}{$\ST\geq1300$}        & $16.8\pm     3.8$                  & $       20$        & $18 \pm 3.5$                               \\
			\hline
			47  & \multirow{3}{*}{\PhoHTG{}}                & \multirow{3}{*}{$\phopT\geq100$}       & \multirow{3}{*}{$\HTG\geq2000$}           & \multicolumn{1}{c|}{$350\leq\ptmiss<450$} & $5.7\pm     2.6$                   & $         4$        &   $6.5 \pm 1.9$                            \\
			48  &                                           &                                        &                                           & \multicolumn{1}{c|}{$450\leq\ptmiss<600$} & $2.7\pm     0.9$                   & $        10$         &  $4.1 \pm 1.2$                            \\
			49  &                                           &                                        &                                           & \multicolumn{1}{c|}{$\ptmiss\geq600$}     & $2.5\pm       1.0$                 & $         4$          &  $3.1 \pm 1.5$                           \\
		\end{tabular}
	}
\end{table*}

The results of the combination are interpreted in terms of the GGM scenario and the simplified models introduced in Section~\ref{sec:ggm}. The 95\% confidence level (\CL) upper limits on the SUSY cross sections are calculated with the \CLs criterion~\cite{Junk:1999kv,Read} using the LHC-style profile likelihood ratio as a test statistic~\cite{LHCCLs} evaluated in the asymptotic approximation~\cite{Cowan:2010js}. Log-normal nuisance parameters are used to describe the systematic uncertainties, which follow the treatment used in the initial searches. The systematic uncertainties on the cross-section for rare background processes as well as the uncertainties assigned to the electron-to-photon misidentification are treated as fully correlated between all four categories. While the first of these uncertainties is estimated to be 50\% in all four categories, the latter uncertainty ranges from 8 to 50\% depending on the category and \phopT{}. The uncertainties in the prediction of vector boson production in association with photons in the \PhoST{} and \PhoLep{} categories, which can be as large as 20\%, are treated as fully correlated, since similar prediction methods are used. In addition, the following sources of uncertainty on the simulation affect the background estimations and signal acceptance: photon identification and isolation efficiency, simulation of pileup, modeling of initial state radiation, determination of the integrated luminosity and jet energy scale. These uncertainties are also treated as fully correlated across search bins. Furthermore, all systematic uncertainties in the signal acceptance, which are mainly dominated by the fast simulation uncertainty~(up to 36\%) in \ptmiss{}, are assumed to be fully correlated among the four categories.

\begin{figure*}[h!]
	\centering
	\includegraphics[width=0.49\textwidth]{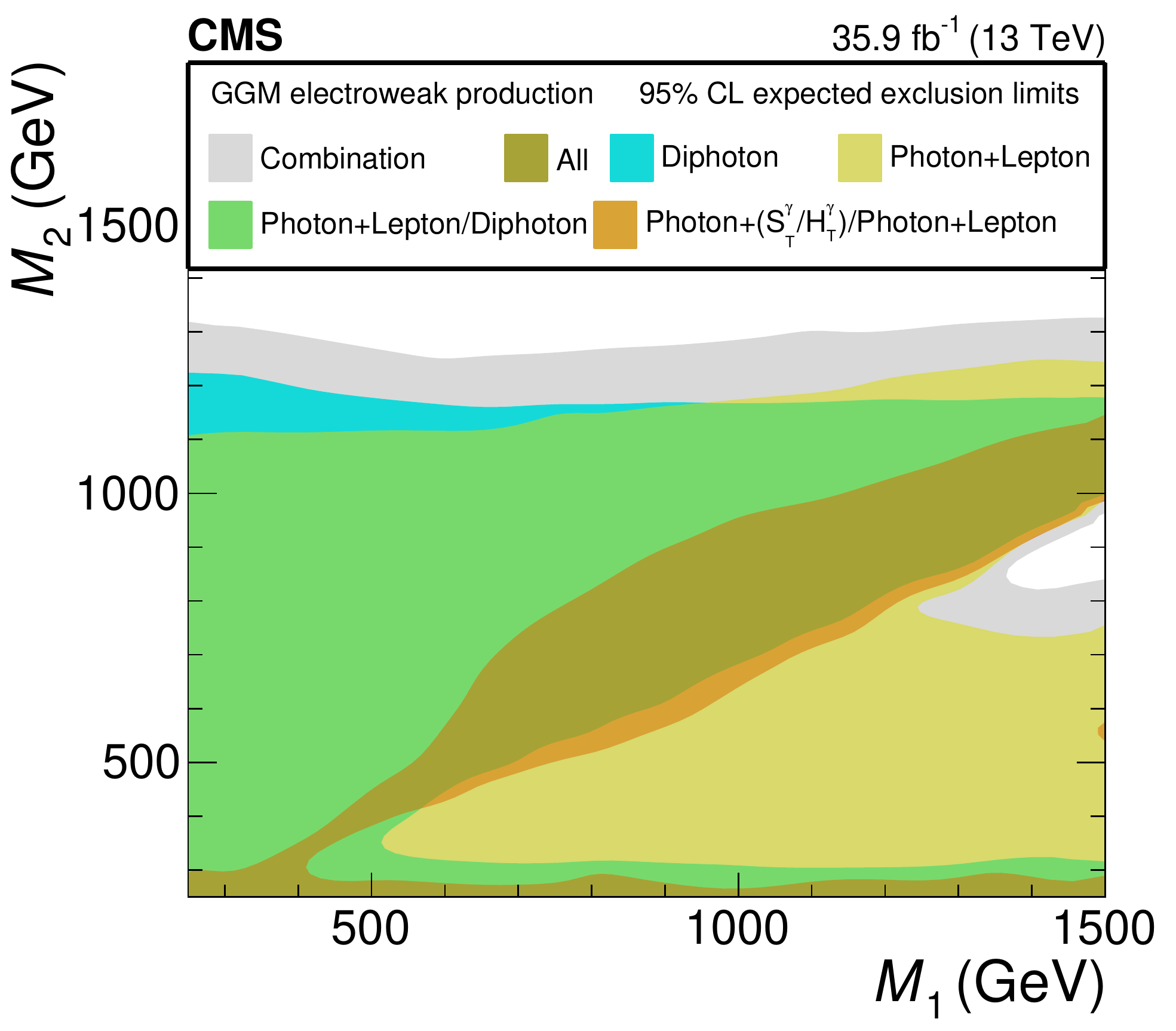}
	\includegraphics[width=0.49\textwidth]{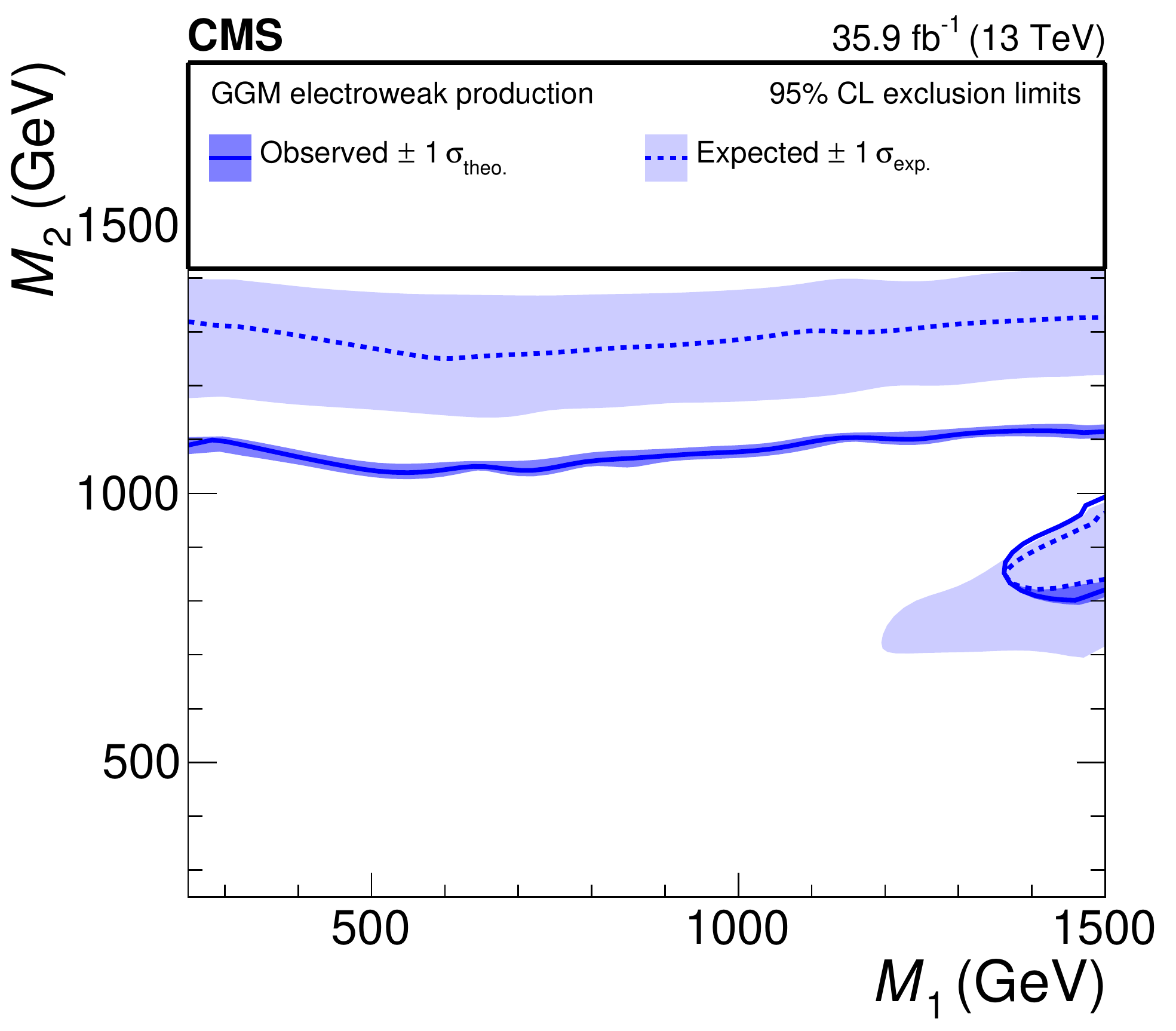}
	\includegraphics[width=0.49\textwidth]{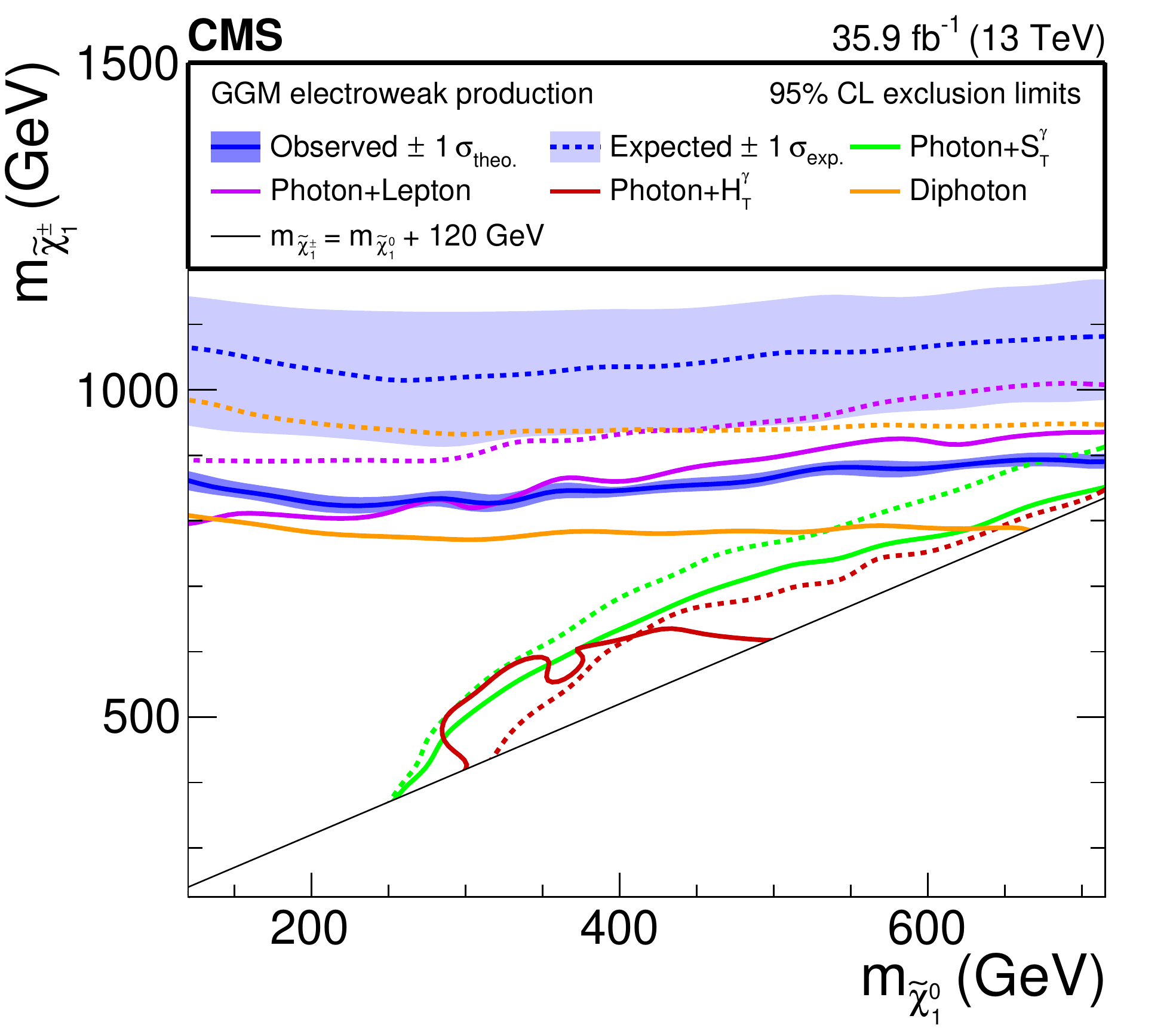}
	\caption{The 95\% \CL exclusion limits for the GGM scenario in terms of the GGM model parameters~(upper) and the physical neutralino and chargino masses~(lower). The upper left panel shows the expected exclusion limits, where the area denoted as ``all'' is excluded by all four individual categories. The upper right panel shows both the corresponding observed~(full lines) and expected~(dotted lines) exclusion limits for the combination in terms of the GGM model parameters. The lower panel shows the observed and the expected exclusion limits for the physical mass plane, where the phase space between the colored lines and the black line is excluded. In the physical mass plane only signal points with a mass difference above 120\GeV are shown to enable a precise projection of the physical masses from the GGM model parameters. The band around the expected limit of the combination indicates the region containing 68\% of the distribution of limits expected under the background-only hypothesis. The band around the observed limit of the combination shows the spread in the observed limit from variation of the signal cross sections within their theoretical uncertainties.}
	\label{fig:resultsGGMM1M2}
\end{figure*}

Results for the GGM scenario are presented in the parameters that are scanned~(\mOne and \mTwo) and in terms of physical mass parameters for the chargino and neutralino. Figure~\ref{fig:resultsGGMM1M2} (upper left) shows the combined expected exclusion limits at 95\% \CL for the GGM scenario, where the combination excludes almost all signal points up to $\mTwo=1300\GeV$ across the full range of \mOne. The figure indicates which category is able to exclude a particular signal point. The grey areas labeled as ''combination'' show the phase space where only the combination of the categories is expected to exclude the signal points at 95\% \CL. The area at large \mOne{}~values, which is only covered by the \PhoLep{} category, corresponds to signal points with a wino-like NLSP reducing the probability of a second high-energy photon in the event. Figure~\ref{fig:resultsGGMM1M2} (upper right) shows both the observed and expected exclusion for the combination in the GGM model parameters. Figure~\ref{fig:resultsGGMM1M2} (lower) shows the observed and expected exclusion limits as a function of the physical masses of the lightest chargino and the lightest neutralino. The exclusion limits of the \Diphoton{} and \PhoLep{} categories are nearly independent of the neutralino mass since these categories have lower \ptmiss{} requirements. The higher \ptmiss{} regions used in the \PhoST{} and \PhoHTG{} categories mainly contribute closer to the mass diagonal at higher neutralino masses. The combination exceeds the sensitivity of the individual searches by around $100\GeV$ with respect to the wino mass parameter~\mTwo{}, which translates to an expected gain of up to 100\GeV for the lightest chargino mass limit. For low neutralino masses, the combination is able to improve the observed limit on the chargino mass by up to 30\GeV. For higher chargino masses, the combination does not improve the current best observed limit mainly because the \Diphoton{} category, which shows an observed excess of about two sigma above the expectation, has large sensitivity in this phase space along with the \PhoLep{} category. Figure~\ref{fig:resultsGGMM1M2} also shows that at higher neutralino masses the expected exclusion limits from the \Diphoton{} and \PhoLep{} categories cross as the branching fraction from photons decreases and the branching fraction to Z bosons increases as shown in Fig.~\ref{fig:BF}.

\begin{figure}[h!]
	\centering
	\includegraphics[width=0.49\textwidth]{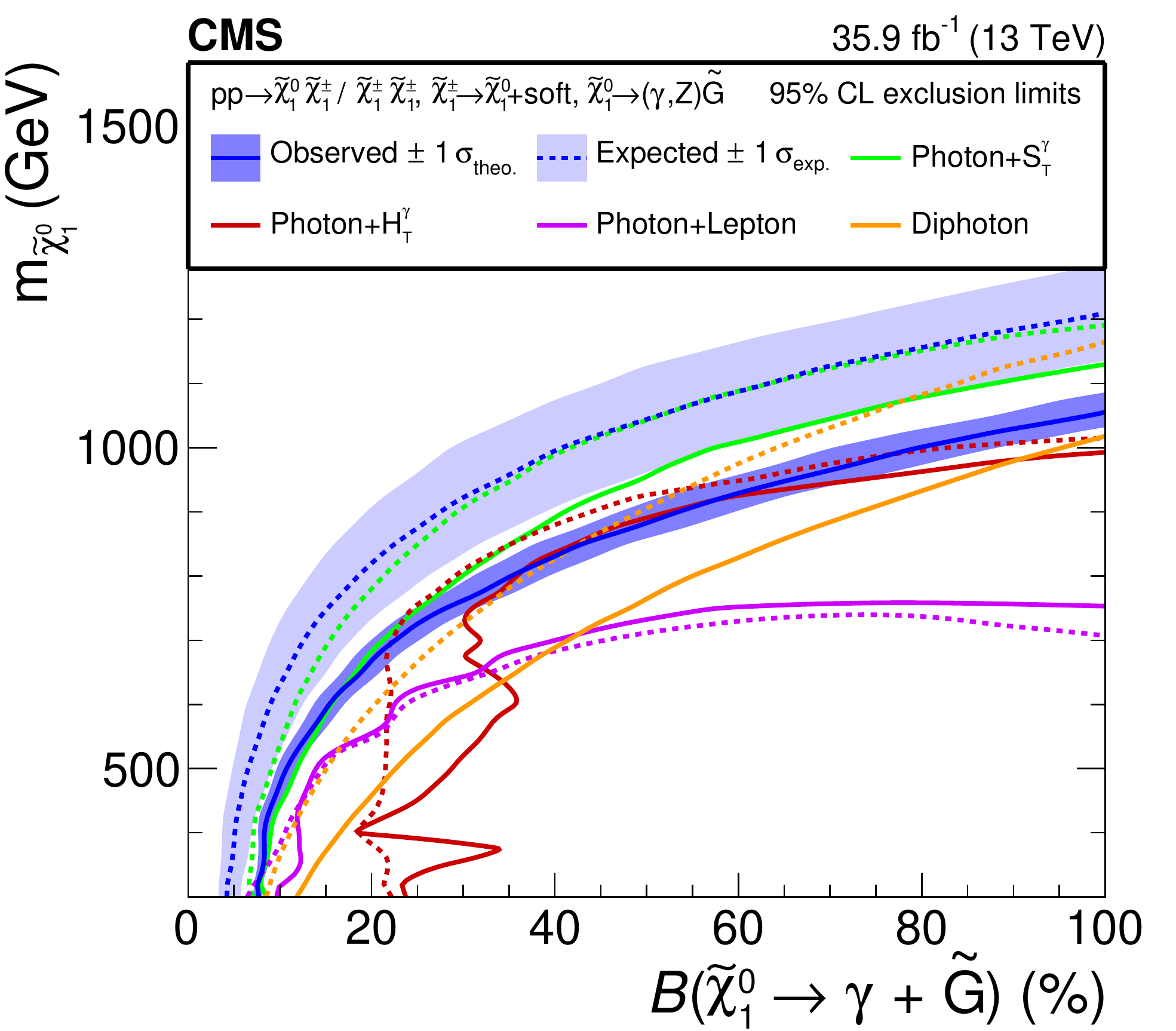}
	\includegraphics[width=0.49\textwidth]{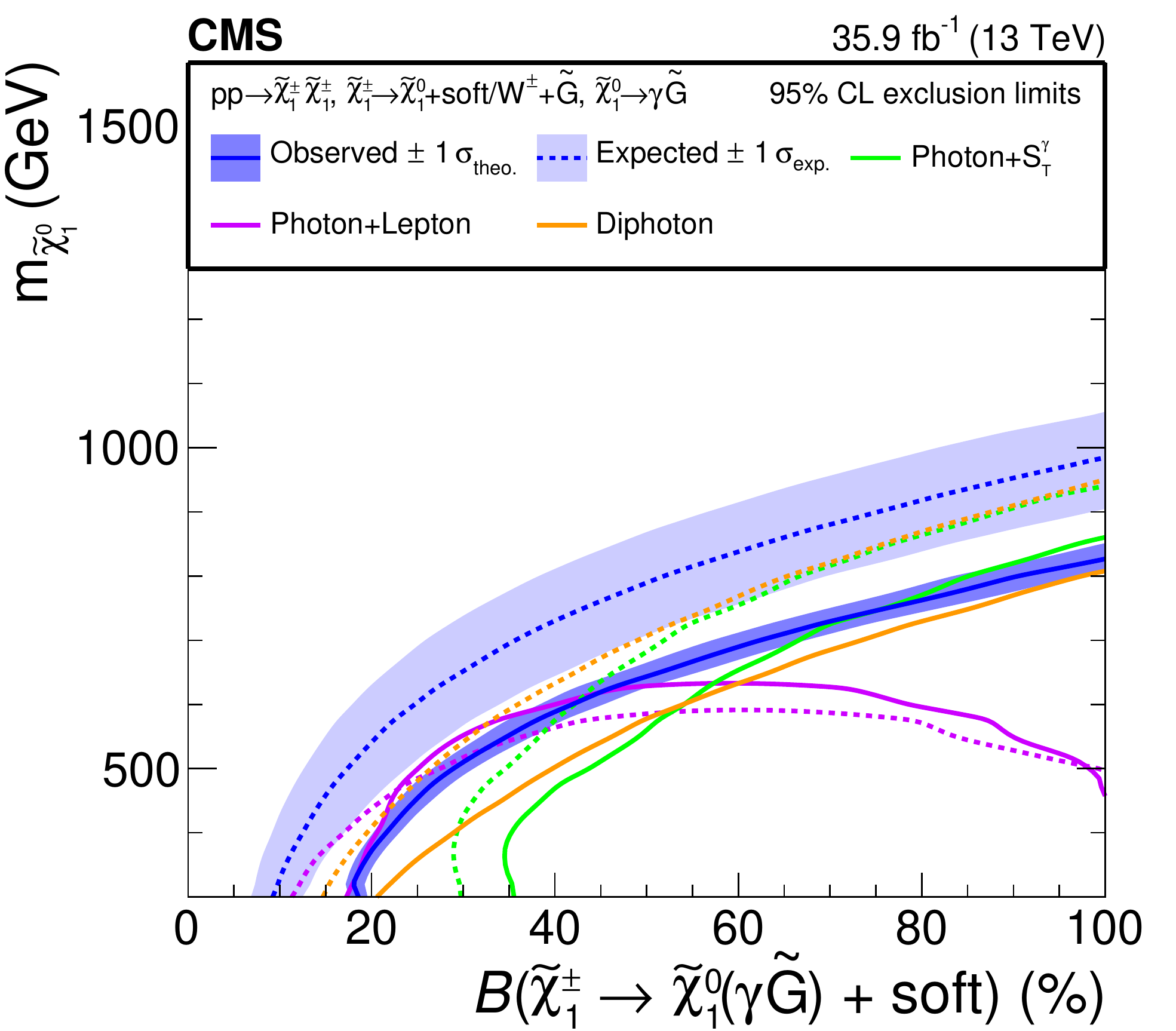}
	\caption{The combined $95\%$ \CL NLSP mass exclusion limits for EW SMS production above 300 GeV. For the neutralino branching fraction scenario (\cmsLeft), the limit is shown as a function of the branching fraction $\chizz\to \gamma + \sGra$, the other decay channel being $\chizz\to \PZ + \sGra$. For the chargino branching fraction scenario (\cmsRight), the limit is shown as a function of the branching fraction $\chipmpm\to \chizz(\gamma\sGra) + \text{soft}$, the other decay channel being $\chipmpm\to \PW + \sGra$. The full lines represent the observed and the dashed lines the expected exclusion limits, where the phase space below the lines is excluded. The band around the expected limit of the combination indicates the region containing 68\% of the distribution of limits expected under the background-only hypothesis. The band around the observed limit of the combination shows the spread in the observed limit from variation of the signal cross sections within their theoretical uncertainties.}
	\label{fig:resultsSMSTChi}
\end{figure}

Figure~\ref{fig:resultsSMSTChi} shows the NLSP mass exclusion limits at 95\% \CL for simplified topologies in EW production scenarios with varying branching fractions of the neutralino~(\cmsLeft) and chargino~(\cmsRight) decay. Here, the \PhoST{} category provides the highest sensitivity along with the \Diphoton{} category. Smaller contributions arise from the \PhoLep{} category. The sensitivity of the \PhoLep{} category to scenarios with large branching fractions of the decay $\chizz\to \gamma + \sGra$ especially arises from events where one photon is misidentified as a lepton. For the neutralino branching fraction scenario, which probes the $\chipmpm\chizz$ and $\chipmpm\chipmmp$ production, the combined expected exclusion limits for NLSP masses ranges from 1200\GeV for a branching fraction of 100\% for the decay $\chizz\to \gamma + \sGra$ to 1000\GeV for 50\%. For smaller branching fractions, the sensitivity for all categories drops since the probability of a final state with at least one photon decreases. This combined exclusion limit almost coincides with the exclusion limit based on the \PhoST{} category. In case of the chargino branching fraction scenario only $\chipmpm\chipmmp$ is produced, leading to a smaller signal cross section. Here, an expected limit on the NLSP mass of up to $1000\GeV$ can be achieved for high branching fractions for the decay $\chipmpm\to \chizz(\gamma\sGra) + \text{soft}$. The largest gain in sensitivity from the combination is found at a branching fraction of 40\%, where the sensitivity of \PhoST{}, \PhoLep{}, and \Diphoton{} categories is of the same order. The \PhoHTG{} category shows no exclusion power for this scenario. Observed gaugino mass limits are set up to 1050 and 825\GeV in the neutralino and the chargino branching fraction scenarios, respectively.

\begin{figure}[h!]
	\centering
	\includegraphics[width=0.49\textwidth]{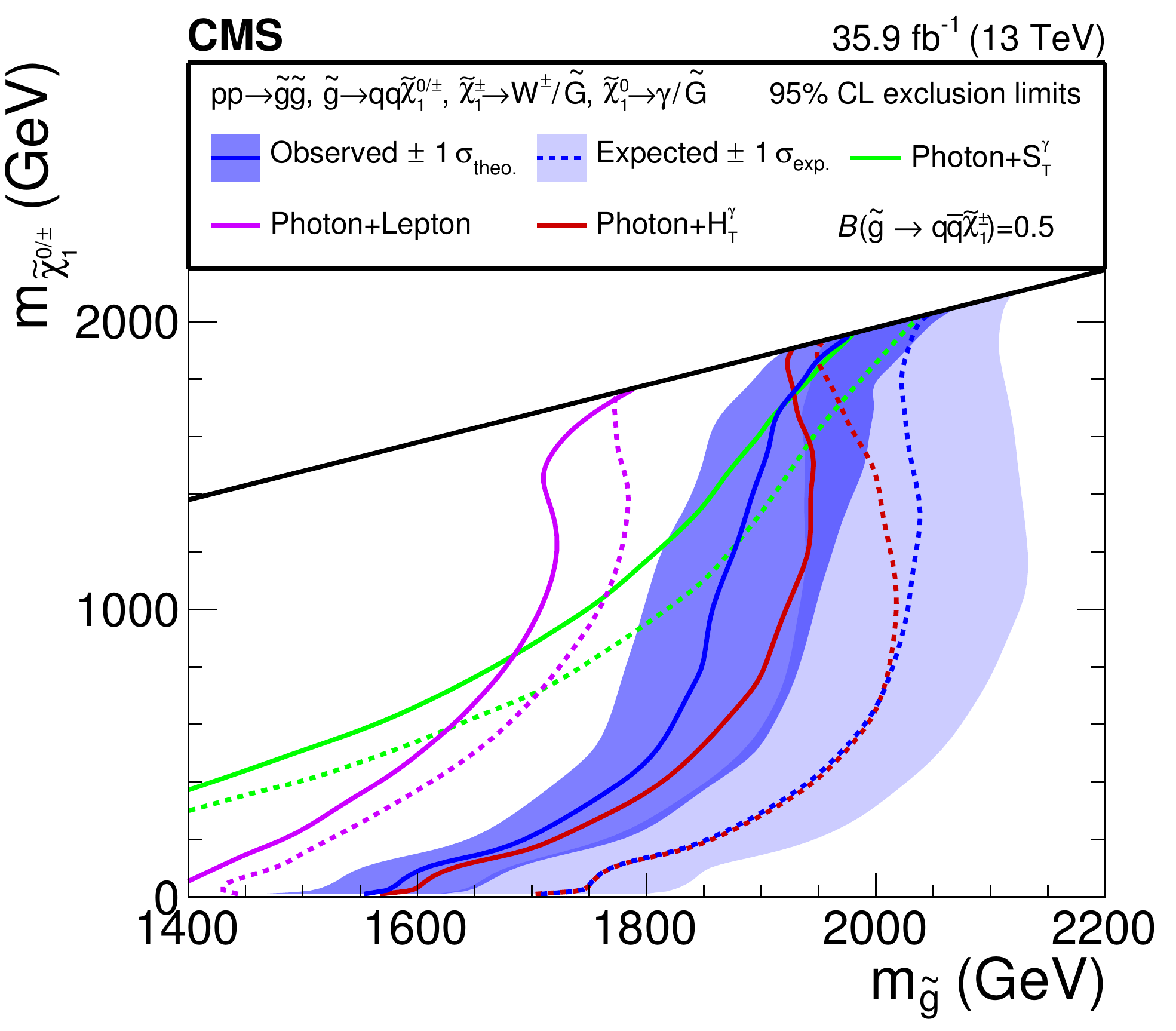}
	\includegraphics[width=0.49\textwidth]{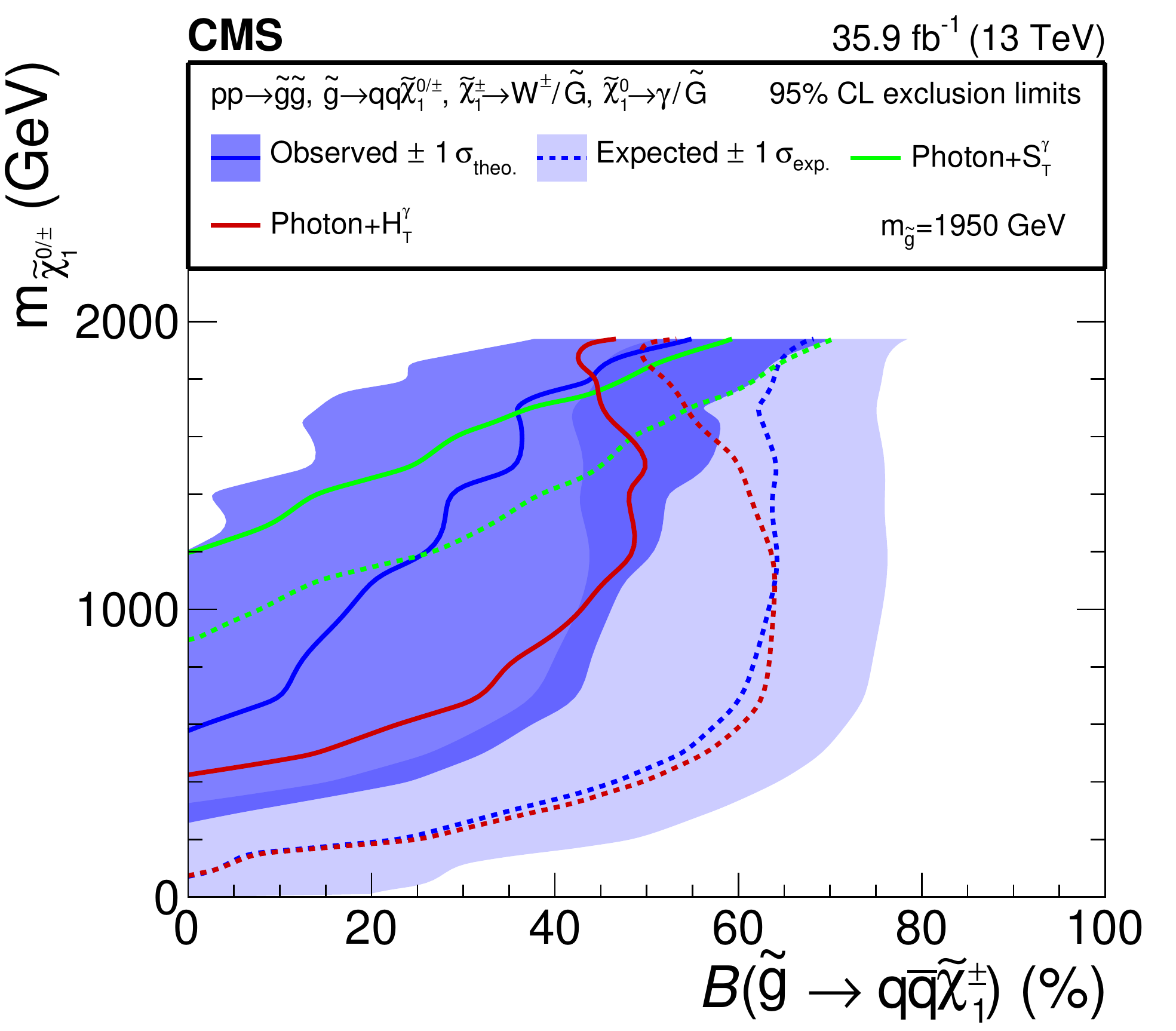}
	\caption{The $95\%$ \CL exclusion limits for the nominal gluino scenario~(\cmsLeft) assuming equal probabilities of 50\% for the gluino decay to $\PQq\PAQq\chipmpm$ and $\PQq\PAQq\chizz$. For the gluino branching fraction scenario~(\cmsRight) the ratio of the probabilities for both decays are scanned and the gluino mass is fixed to 1950\GeV. The \PhoLep{} category shows no exclusion power for the latter scenario. The full lines represent the observed and the dashed lines the expected exclusion limits, where the phase space below the lines is excluded. The band around the expected limit of the combination indicates the region containing 68\% of the distribution of limits expected under the background-only hypothesis. The band around the observed limit of the combination shows the spread in the observed limit from variation of the signal cross sections within their theoretical uncertainties.}
	\label{fig:resultsSMST5Wg}
\end{figure}

The results from simplified topologies in strong production of gluinos are shown in Fig.~\ref{fig:resultsSMST5Wg}. For these topologies the sensitivity of \Diphoton{} category is reduced and therefore not included in the combination, which allows for a removal of the diphoton veto discussed in Section~\ref{sec:EventSelection} and mainly increases the sensitivity of the \PhoHTG{} category. Table~\ref{tab:yields_strong} shows the data and the background prediction yields without the diphoton veto. In case of the nominal gluino scenario, introduced in Section~\ref{sec:ggm}, the combination shows an optimal expected exclusion compared to the different individual categories across a broad region of the mass parameter space. For NLSP masses below $1000\GeV$, the sensitivity of the combination is dominated by the \PhoHTG{} category, which mainly targets signal events with large hadronic activity. However, at NLSP masses above 1700\GeV, the \PhoST{} category, which benefits from the smaller hadronic activity close to the mass diagonal, provides the highest sensitivity. The \PhoLep{} category selects events where the \PW~boson decays leptonically, leading to a reduced sensitivity compared to the inclusive categories. The largest improvement of the combination is achieved in the phase space where the sensitivity of both inclusive categories is of the same order. Here, the expected limit on the gluino mass is improved by 50\GeV. The \cmsRight plot of Fig.~\ref{fig:resultsSMST5Wg} shows the limits for the same SMS topology with a fixed gluino mass of $1950\GeV$ but with the gluino branching fraction varied between its decays to $\PQq\PAQq\chipmpm$ and $\PQq\PAQq\chizz$. Compared to the nominal gluino scenario similar behavior in the two inclusive categories is found.

\begin{table*}[htb!]
	\topcaption{Predicted pre-fit background yields, where the values are not constrained by the likelihood fit, the observed number of events in data, and the post-fit background yields after the constraint from the likelihood fit
 for all search bins used in the combination. In addition the range covered by each individual bin is shown. For these yields, the \Diphoton{} category is not included and the \Diphoton{} veto is removed to increase the sensitivity of the \PhoHTG{} category to strong production of gluinos. }
	\label{tab:yields_strong}
	\centering
	\renewcommand{\arraystretch}{1.1}
	\cmsTableWide{
		\centering
		\begin{tabular}{c|c|c|c|c|c|c|c}
			Bin & Category                                  & \multicolumn{3}{c|}{Ranges (\GeVns)}   & Total bkg.                         & Data      & Post-fit Total bkg.                                          \\ \hline
			7   & \multirow{18}{*}{\PhoLep{} ($\mu\gamma$)} & \multirow{9}{*}{$35\leq\phopT<200$}    & \multirow{3}{*}{$120\leq\ptmiss<200$} & $\HT<100$     & $317\pm    50$      & $       309$ &  $320 \pm 18$ \\
			8   &                                           &                                        &                                    & $100\leq\HT<400$ & $470\pm    98$      & $       501$  & $489 \pm 31$\\
			9   &                                           &                                        &                                    & $\HT\geq400$     & $100\pm      27$    & $        86$ & $99 \pm 7$\\
			\cline{4-8}
			10  &                                           &                                        & \multirow{3}{*}{$200\leq\ptmiss<400$} & $\HT<100$     & $26.3\pm     5.3$   & $        33$ & $30.7 \pm 3.4$ \\
			11  &                                           &                                        &                                    & $100\leq\HT<400$ & $61\pm    14$       & $        65$ & $63 \pm 5$\\
			12  &                                           &                                        &                                    & $\HT\geq400$     & $45\pm    14$       & $        45$ & $46 \pm 4$\\
			\cline{4-8}
			13  &                                           &                                        & \multirow{3}{*}{$\ptmiss\geq400$}     & $\HT<100$     & $1.2\pm     0.4$    & $         1$ & $1.3 \pm 0.4$\\
			14  &                                           &                                        &                                    & $100\leq\HT<400$ & $2.4\pm     1.1$    & $         1$ & $2.1 \pm 0.7$\\
			15  &                                           &                                        &                                    & $\HT\geq400$     & $5.3\pm       2.0$  & $         5$ & $5.4 \pm 1.0$\\
			\cline{3-8}
			16  &                                           & \multirow{9}{*}{$\phopT\geq200$}          & \multirow{3}{*}{$120\leq\ptmiss<200$} & $\HT<100$     & $6.3\pm     2.4$    & $        12$ & $9.1 \pm 1.6$  \\
			17  &                                           &                                        &                                    & $100\leq\HT<400$ & $21.1\pm     7.2$   & $        24$ & $23.3 \pm 2.4$\\
			18  &                                           &                                        &                                    & $\HT\geq400$     & $15.3\pm     4.8$   & $        20$ & $17.3 \pm 1.8$\\
			\cline{4-8}
			19  &                                           &                                        & \multirow{3}{*}{$200\leq\ptmiss<400$} & $\HT<100$     & $4.8\pm     1.8$    & $         4$ & $6.9 \pm 1.1$\\
			20  &                                           &                                        &                                    & $100\leq\HT<400$ & $8.3\pm     3.2$    & $        13$ & $9.1 \pm 1.1$\\
			21  &                                           &                                        &                                    & $\HT\geq400$     & $5.4\pm       2.0$  & $         7$ & $6.3 \pm 0.9$\\
			\cline{4-8}
			22  &                                           &                                        & \multirow{3}{*}{$\ptmiss\geq400$}     & $\HT<100$     & $0.7\pm     0.4$    & $         1$ & $1.3 \pm 0.3$\\
			23  &                                           &                                        &                                    & $100\leq\HT<400$ & $0.6\pm     0.2$    & $         1$ & $0.8 \pm 0.2$\\
			24  &                                           &                                        &                                    & $\HT\geq400$     & $0.5\pm     0.2$    & $         0$ & $0.6 \pm 0.1$\\
			\hline
			25  & \multirow{18}{*}{\PhoLep{} $(\Pe\gamma)$}   & \multirow{9}{*}{$35\leq\phopT<200$}          & \multirow{3}{*}{$120\leq\ptmiss<200$} & $\HT<100$     & $166\pm    22$      & $       154$ & $167 \pm 11$\\
			26  &                                           &                                        &                                    & $100\leq\HT<400$ & $261\pm    53$      & $       276$ & $269 \pm 18$\\
			27  &                                           &                                        &                                    & $\HT\geq400$     & $80\pm    21$       & $        67$ & $79 \pm 7$\\
			\cline{4-8}
			28  &                                           &                                        & \multirow{3}{*}{$200\leq\ptmiss<400$} & $\HT<100$     & $17.3\pm     3.2$   & $        32$ & $21.2 \pm 2.9$\\
			29  &                                           &                                        &                                    & $100\leq\HT<400$ & $51\pm    12$       & $        46$ & $51.1 \pm 4.1$\\
			30  &                                           &                                        &                                    & $\HT\geq400$     & $28.8\pm       9.0$ & $        32$ & $29.5 \pm 2.9$\\
			\cline{4-8}
			31  &                                           &                                        & \multirow{3}{*}{$\ptmiss\geq400$}     & $\HT<100$     & $1.3\pm     0.5$    & $         1$ & $1.5 \pm 0.4$ \\
			32  &                                           &                                        &                                    & $100\leq\HT<400$ & $1.2\pm     0.5$    & $         1$ & $1.2 \pm 0.4$\\
			33  &                                           &                                        &                                    & $\HT\geq400$     & $2.8\pm     0.8$    & $         4$ & $3.3 \pm 0.6$\\
			\cline{3-8}
			34  &                                           & \multirow{9}{*}{$\phopT\geq200$}          & \multirow{3}{*}{$120\leq\ptmiss<200$} & $\HT<100$     & $6.3\pm     2.1$    & $        10$ & $8.3 \pm 1.3$  \\
			35  &                                           &                                        &                                    & $100\leq\HT<400$ & $21.8\pm     7.1$   & $        22$ & $23.4 \pm 2.4$\\
			36  &                                           &                                        &                                    & $\HT\geq400$     & $11.7\pm     3.7$   & $        15$ & $13.0 \pm 1.4$\\
			\cline{4-8}
			37  &                                           &                                        & \multirow{3}{*}{$200\leq\ptmiss<400$} & $\HT<100$     & $4.4\pm     1.6$    & $         7$ & $5.9 \pm 1.0$ \\
			38  &                                           &                                        &                                    & $100\leq\HT<400$ & $8.9\pm     3.3$    & $         9$ & $9.3 \pm 1.2$\\
			39  &                                           &                                        &                                    & $\HT\geq400$     & $5.1\pm     1.8$    & $         4$ & $5.6 \pm 0.7$\\
			\cline{4-8}
			40  &                                           &                                        & \multirow{3}{*}{$\ptmiss\geq400$}     & $\HT<100$     & $0.4\pm     0.2$    & $         0$ & $0.7 \pm 0.2$\\
			41  &                                           &                                        &                                    & $100\leq\HT<400$ & $0.5\pm     0.2$    & $         1$ & $0.7 \pm 0.2$\\
			42  &                                           &                                        &                                    & $\HT\geq400$     & $0.8\pm     0.5$    & $         3$ & $1.1 \pm 0.4$\\
			\hline
			43  & \multirow{4}{*}{\PhoST{}}                 & \multirow{4}{*}{$\phopT\geq180$}       & \multirow{4}{*}{$\HTG\leq2000$}           & \multicolumn{1}{c|}{$600\leq\ST<800$}     & $261\pm    30$                     & $       275$    &                              $273 \pm 22$     \\
			44  &                                           &                                        &                                           & \multicolumn{1}{c|}{$800\leq\ST<1000$}    & $97\pm    14$                      & $       98$      & $100 \pm 9$                                 \\
			45  &                                           &                                        &                                           & \multicolumn{1}{c|}{$1000\leq\ST<1300$}   & $50.2\pm     8.0$                  & $       59$       & $53.8 \pm 6.7$                                \\
			46  &                                           &                                        &                                           & \multicolumn{1}{c|}{$\ST\geq1300$}        & $16.9\pm     3.8$                  & $       20$        & $18.0 \pm 3.1$                               \\
			\hline
			47  & \multirow{3}{*}{\PhoHTG{}}                & \multirow{3}{*}{$\phopT\geq100$}       & \multirow{3}{*}{$\HTG\geq2000$}           & \multicolumn{1}{c|}{$350\leq\ptmiss<450$} & $5.7\pm     2.6$                   & $         5$        & $6.9 \pm 2.2$                              \\
			48  &                                           &                                        &                                           & \multicolumn{1}{c|}{$450\leq\ptmiss<600$} & $2.7\pm     0.9$                   & $        10$         &$4.1 \pm 1.4$                              \\
			49  &                                           &                                        &                                           & \multicolumn{1}{c|}{$\ptmiss\geq600$}     & $2.6\pm       1.0$                 & $         4$          &    $3.1 \pm 1.5$                        \\
		\end{tabular}
	}
\end{table*}

In most of the simplified topologies, the combination of the different categories outperforms the individual searches with respect to the expected limit. The \cmsRight plot of Fig.~\ref{fig:resultsSMST5Wg} shows a slight degradation of the expected limit at medium branching fractions for the combination compared to the \PhoHTG{} category. This is caused by the removal of the events with moderate \HTG{} and lepton events from the \PhoHTG{} category, as explained in Section~\ref{sec:EventSelection}. This strategy is motivated by optimizing the sensitivity to the GGM scenario shown in Fig.~\ref{fig:resultsGGMM1M2}. Small excesses in data with respect to the background prediction are found in each of the four categories, which give rise to differences in the observed and expected limits. As a result, only small improvements are made in the observed limits compared to the individual searches in all interpretations.

\section{Summary}
\label{sec:sum}

A combination of four different searches for general gauge-mediated~(GGM) supersymmetry~(SUSY) in final states with photons and a large transverse momentum imbalance was performed. Based on the event selection of the individual searches, four event categories were defined. Overlaps between the categories were removed by additional vetoes designed to maximize the sensitivity of the combination. Using data recorded with the CMS detector at the LHC at a center-of-mass energy of 13\TeV, and corresponding to an integrated luminosity of 35.9\fbinv, the combination improves the expected sensitivity of the searches described in Ref.~\cite{photonst, photonht, photonlep, diphoton}.

The results are interpreted in the context of GGM SUSY and in simplified models. The sensitivity of the combination is also interpreted across a range of branching fractions, allowing for generalization to a wide range of SUSY scenarios. The results of the GGM scenario are expressed as limits on the physical mass parameters. Here, chargino masses up to~890\,(1080)\GeV are excluded by the observed~(expected)~limit across the tested neutralino mass spectrum, which ranges from $120$ to 720\GeV. In electroweak production models, limits for neutralino masses are set up to 1050\,(1200)\GeV for combined $\chipmpm\chizz$ and $\chipmpm\chipmmp$ production, while for pure $\chipmpm\chipmmp$ production these limits are reduced to 825\,(1000)\GeV. For a strong production scenario based on gluino pair production, the highest excluded gluino mass is at $1975\,(2050)\GeV$. The combination improves on the expected limits on neutralino and chargino masses by up to 100\GeV, while the expected limit on the gluino mass is increased by 50\GeV compared to the individual searches.

\begin{acknowledgments}
We wish to acknowledge the help of Simon Knapen, David Shih, and Diego Redigolo, who  provided us with the signal model used in this analysis.
	
We congratulate our colleagues in the CERN accelerator departments for the excellent performance of the LHC and thank the technical and administrative staffs at CERN and at other CMS institutes for their contributions to the success of the CMS effort. In addition, we gratefully acknowledge the computing centres and personnel of the Worldwide LHC Computing Grid for delivering so effectively the computing infrastructure essential to our analyses. Finally, we acknowledge the enduring support for the construction and operation of the LHC and the CMS detector provided by the following funding agencies: BMBWF and FWF (Austria); FNRS and FWO (Belgium); CNPq, CAPES, FAPERJ, FAPERGS, and FAPESP (Brazil); MES (Bulgaria); CERN; CAS, MoST, and NSFC (China); COLCIENCIAS (Colombia); MSES and CSF (Croatia); RPF (Cyprus); SENESCYT (Ecuador); MoER, ERC IUT, PUT and ERDF (Estonia); Academy of Finland, MEC, and HIP (Finland); CEA and CNRS/IN2P3 (France); BMBF, DFG, and HGF (Germany); GSRT (Greece); NKFIA (Hungary); DAE and DST (India); IPM (Iran); SFI (Ireland); INFN (Italy); MSIP and NRF (Republic of Korea); MES (Latvia); LAS (Lithuania); MOE and UM (Malaysia); BUAP, CINVESTAV, CONACYT, LNS, SEP, and UASLP-FAI (Mexico); MOS (Montenegro); MBIE (New Zealand); PAEC (Pakistan); MSHE and NSC (Poland); FCT (Portugal); JINR (Dubna); MON, RosAtom, RAS, RFBR, and NRC KI (Russia); MESTD (Serbia); SEIDI, CPAN, PCTI, and FEDER (Spain); MOSTR (Sri Lanka); Swiss Funding Agencies (Switzerland); MST (Taipei); ThEPCenter, IPST, STAR, and NSTDA (Thailand); TUBITAK and TAEK (Turkey); NASU and SFFR (Ukraine); STFC (United Kingdom); DOE and NSF (USA).
\end{acknowledgments}

\bibliography{auto_generated}
\cleardoublepage \appendix\section{The CMS Collaboration \label{app:collab}}\begin{sloppypar}\hyphenpenalty=5000\widowpenalty=500\clubpenalty=5000\vskip\cmsinstskip
\textbf{Yerevan Physics Institute, Yerevan, Armenia}\\*[0pt]
A.M.~Sirunyan$^{\textrm{\dag}}$, A.~Tumasyan
\vskip\cmsinstskip
\textbf{Institut f\"{u}r Hochenergiephysik, Wien, Austria}\\*[0pt]
W.~Adam, F.~Ambrogi, T.~Bergauer, J.~Brandstetter, M.~Dragicevic, J.~Er\"{o}, A.~Escalante~Del~Valle, M.~Flechl, R.~Fr\"{u}hwirth\cmsAuthorMark{1}, M.~Jeitler\cmsAuthorMark{1}, N.~Krammer, I.~Kr\"{a}tschmer, D.~Liko, T.~Madlener, I.~Mikulec, N.~Rad, J.~Schieck\cmsAuthorMark{1}, R.~Sch\"{o}fbeck, M.~Spanring, D.~Spitzbart, W.~Waltenberger, C.-E.~Wulz\cmsAuthorMark{1}, M.~Zarucki
% ...
\vskip\cmsinstskip
\dag: Deceased\\
1:  Also at Vienna University of Technology, Vienna, Austria\\
% ...
\end{sloppypar}
\end{document}